\documentclass[useAMS,usenatbib]{mn2e}
\usepackage{aas_macros,graphicx,times,amsmath,amssymb,color,longtable}

\title[A 0.42-s ULX  pulsar in NGC\,7793]{Discovery of a 0.42-s pulsar in the ultraluminous X-ray source NGC\,7793 P13}
\author[G.L.~Israel et al.] {G.~L.~Israel,$^{1}$\thanks{E-mail: gianluca@oa-roma.inaf.it} 
A. Papitto,$^{1}$  P. Esposito,$^{2,4}$ L. Stella,$^{1}$ L. Zampieri,$^{3}$ A.~Belfiore,$^{4}$ \newauthor G.~A.~Rodr\'iguez Castillo,$^{1}$  A.~De~Luca,$^{4}$ A.~Tiengo,$^{5,4,6}$ F.~Haberl,$^{7}$ J. Greiner,$^{7}$ \newauthor R.~Salvaterra,$^{4}$ S.~Sandrelli$^8$ and G. Lisini$^{5}$
\smallskip\\
$^1$ INAF--Osservatorio Astronomico di Roma, via Frascati 33, I-00040 Monteporzio Catone, Italy\\
$^{2}$ Anton Pannekoek Institute for Astronomy, University of Amsterdam,
Postbus 94249, NL-1090-GE Amsterdam, The Netherlands\\
$^{3}$ INAF--Osservatorio Astronomico di Padova, INAF, vicolo dell'Osservatorio 5, I-35122 Padova, Italy\\
$^4$ INAF--Istituto di Astrofisica Spaziale e Fisica Cosmica - Milano, via E. Bassini 15, I-20133 Milano, Italy\\
$^5$ Scuola Universitaria Superiore IUSS Pavia, piazza della Vittoria 15, I-27100 Pavia, Italy \\
$^6$ INFN-–Istituto Nazionale di Fisica Nucleare, Sezione di Pavia, via A. Bassi 6, I-27100 Pavia, Italy\\
$^7$Max-Planck-Institut f\"ur extraterrestrische Physik, Giessenbachstra{\ss}e, D-85748 Garching, Germany\\
$^8$INAF--Osservatorio Astronomico di Brera, via Brera 28, I-20121 Milano, Italy\\
}
\date{Accepted 2016 October 14.  Received 2016 October 13; in original form 2016 September 20} \pagerange{\pageref{firstpage}--\pageref{lastpage}} \pubyear{2017}

\def\LaTeX{L\kern-.36em\raise.3ex\hbox{a}\kern-.15em
    T\kern-.1667em\lower.7ex\hbox{E}\kern-.125emX}
\def\xmm {\emph{XMM--Newton}}
\def\nustar {\emph{NuSTAR}}
\def\cxo {\emph{Chandra}}
\def\ein {\emph{Einstein}}
\def\swift {\emph{Swift}}

\def\rst {\emph{ROSAT}}

\def\src {\mbox{NGC\,7793\,P13}}
\def\flux {\mbox{erg cm$^{-2}$ s$^{-1}$}}
\def\lum {\mbox{erg s$^{-1}$}}

\def\srcc{\mbox{CXOU J235750.9--323726}}
\def\srcx{\mbox{XMMU J235751.1--323725}}
\def\rc {\rm}
\def\rcb {\rm}

\begin{document}

\label{firstpage}
\maketitle
\begin{abstract}
\src\ is  a variable (luminosity range {\rc $\sim$ 100}) ultraluminous X-ray source (ULX) proposed to host a stellar-mass black hole of less than 15~M$_{\odot}$ in a binary system with orbital period of 64~d and a 18--23~M$_{\odot}$ B9Ia companion. Within the EXTraS project we discovered pulsations at a period of $\sim$0.42~s in two \xmm\ observations of \src, during which the source was detected at $L_{\mathrm{X}}\sim2.1\times10^{39}$ and $5\times10^{39}$~\lum\ (0.3--10~keV band). These findings unambiguously demonstrate that the compact object in \src\ is a neutron star  accreting at super-Eddington rates. While standard accretion models face difficulties  accounting for the pulsar X-ray luminosity, the presence of a multipolar magnetic field with $B\sim$ few $\times$ 10$^{13}$\,G close to the base of the accretion column appears to be in agreement with the properties of the system. 
\end{abstract}


\section{Introduction}
Ultraluminous X-ray sources (ULXs) are extra-nuclear point-like X-ray objects located in  nearby galaxies with X-ray luminosities exceeding the Eddington limit of $>$10$^{39}$\lum\ for a $\sim$10 M$_{\odot}$ black hole (BH).
Based on their spectral and timing properties, it has been proposed (see \citealt{roberts16} for a recent review) that most ULXs are stellar-remnant black holes (with masses possibly reaching $\sim$100 M$_{\odot}$; \citealt{zampieri09,belczynski10}) accreting at  super-Eddington rates. The salient ULX features revealed by \xmm\ and \nustar\ observations and supporting the scenario of super-Eddington accretion onto BHs is a downturn of the X-ray spectrum at energies of $\sim$5--10~keV and a soft excess at lower energies (\citealt{roberts16} and references therein). Despite evidences in favour of the BH nature of of the compact remnant in ULXs \citep{liu13},  there 
are also two notable exceptions of pulsating ULXs (PULXs) testifying to the presence of accreting neutron stars \citep{bachetti14,israel16b}.
This shows that spectral properties alone are not an unambiguous way for a correct identification of the compact remnant in  ULX \citep{bachetti16}. 

Within the framework of the EXTraS\footnote{See http://www.extras-fp7.eu/.} (Exploring the X-ray Transient and variable Sky; \citealt{deluca15}) project, we searched for coherent periodic signals in the about 290,000 time series of sources, with more than 50 counts, detected by \xmm\ in all EPIC public data.  Among dozens of new X-ray pulsators found so far with periodic signals detected at high confidence ($>$4.5$\sigma$), there is \srcx\ = \srcc, also known as the ULX P13 in NGC\,7793.

\src\ was first observed in 1979 by the \ein\ satellite as a bright, $L_{\mathrm{X}}\sim2\times10^{39}$~\lum\ (in the 0.3--10~keV range), X-ray stellar-like object in the nearby ($D= 3.9$~Mpc; \citealt{karachentsev03}{\rc; this distance is adopted throughout the paper}) reasonably face-on ($i= 53.7\degr$) galaxy NGC\,7793 in the Sculptor group \citep{fabbiano92}. It was also detected by \rst\  in 1992  at  $L_{\mathrm{X}}\sim$$3.5\times10^{39}$~\lum\  (value extrapolated in the 0.3--10~keV band; \citealt{read99}). 
A \cxo\ pointing carried out in 2003 September revealed two sources at the \rst\ position of \src, separated by 2~arcsec, namely \srcc\ and CXOU J235750.9--323728 \citep{pannuti11}. The latter source is thought to be unrelated to \src\ and is more than an order of magnitude less luminous than \src\ itself. Their luminosities are    $\sim$$1.2\times10^{39}$ and $\sim$$6.2\times10^{37}$~\lum, respectively.
The compact object in \src\ is orbiting around a B9Ia spectral-type star of 18--23\, M$_{\odot}$ in a binary system with an orbital period of about 64 days and a moderate eccentricity $e$ of 0.3--0.4 \citep{motch14}. By modelling the strong optical and UV orbital modulation, likely arising from the heating of the donor star, a mass for the suspected BH of less than about 15~M$_{\odot}$ has been inferred for \src\ \citep{motch14}.  

Here we report on the discovery of coherent pulsations at a period of 0.42~s in the  EPIC pn lightcurves of \srcx, with a secular first period derivative of $\dot{P}_{\mathrm{sec}}\sim - 4\times10^{-11}$~s~s$^{-1}$ (Sect.\,\ref{timing}). These findings clearly indicate that \src\ hosts an accreting neutron star (NS) in a binary system and not a  stellar-mass BH as previously assumed. We discuss the nature of this new ultraluminous X-ray pulsar (Sect.\,\ref{discussion}), the third discovered so far, and also the fastest-spinning one.

\section{Observations and data analysis}\label{observations}

\subsection{\xmm\ and \cxo}

The region of \src\ was observed by \xmm\ with the EPIC detectors in full imaging mode (Full Frame). The source position was always off-axis, at  angles  from $\sim$1.2 to 4.1~arcmin. The 0.42-s pulsations  were detected in the pn data only (MOS cameras time resolution $\sim$2.6~s; pn time resolution $\sim$73~ms). The public pn data sets covering the position of \src\  are listed in Table\,\ref{obslog}. During the first pointing, a faint source was detected at a  flux level of $\sim$$1.7(3)\times10^{-14}$~\flux, corresponding to a luminosity of $\sim$$3\times10^{37}$~\lum\ and providing only about 120 events.  
\begin{table}
\centering \caption{Logbook of the \xmm\ observations used in this work.} \label{obslog}
\begin{tabular}{@{}lllrl}
\hline
Obs.\,ID & Start date & {Exp.}  & Off-axis$^{a}$ & Count rate$^{b}$ \\
&  & (ks) & (arcsec) & ( cts s$^{-1}$) \\
\hline
0693760101 & 2012-05-14 & 35  & 77 & $0.007\pm0.002  $\\
0693760401 & 2013-11-25 & 45  & 77 &$0.29\pm0.02$   \\
0748390901 & 2014-12-10 & 47  & 249 &$0.53\pm0.02$   \\
\hline
\end{tabular}
\begin{list}{}{}
\item[$^{a}$] Radial off-axis angle of \src\ from the boresight of the telescope.
\item[$^{b}$] Net source count rate from pn in the 0.3--10~keV energy band, using the extraction regions described in the text.
\end{list}
\end{table}

As part of the EXTraS reduction pipeline, the raw observation data files (ODF) were processed with the Science Analysis Software (\textsc{sas}) v.16. Time intervals with high particle background were filtered. Photon event lists and spectra were extracted in a radius of 32~arcsec for the source, while to estimate the background we used a nearby region with radius of 45~arcsec, far from other sources and avoiding CCD gaps. 
Event times were converted to the barycentre of the Solar system with the \textsc{sas} task \textsc{barycen} using the \cxo\ source position ($\rm RA=23^{h} 57^{m} 50\fs9$, $\rm Dec.=-32^{\circ} 37' 26\farcs6$). Spectra were rebinned so as to obtain a minimum of 30 counts per energy bin, and for each spectrum we generated the response matrix and the ancillary file using the \textsc{sas} tasks \textsc{rmfgen} and \textsc{arfgen}. 

{\rc \src\ was imaged 4 times by \cxo\ between 2003 (one pointing) and 2011 (three pointings) with the Advanced CCD Imaging Spectrometer (time resolution of 3.2~s) at an off-axis angle between 2 and 4.5 arcsec. We used \textsc{wavdetect} (within the \textsc{CIAO} package, v.4.8) for source detection and considered six wavelet detection scales (1, 2, 4, 8, 12 and 16 pixels).  Spectra were obtained in the 0.3-10\,keV energy range using the \textsc{specextract},  circular regions of 4 arcsec for the source, and locally optimized background regions for each observation. 
As a result of this analysis, we detect only one source at the position of \src\ in all observations (at variance with  \citealt{pannuti11}). In the following, we assume that the faint source detected by \cxo\ (2011) and \xmm\ (2012) is indeed \src.}\\

\subsection{\swift}

We analysed the 78 observations of NGC\,7793 performed with the \swift\ X-Ray Telescope (XRT; \citealt{burrows05}) between 2010 August 16 and 2016 August 24, for a total exposure of 210~ks. We processed and analysed the observations performed in photon counting (PC) mode using the standard software (\textsc{HEASoft} v. 6.19) and calibration files (\textsc{caldb} v. 20160609). We extracted source photons in a 20-px radius (equivalent to 47.2~arcsec) around the source position; background was extracted from close-by source-free regions. Photons with grade between 0 and 12 were retained in the analysis. 
\begin{figure}
\centering
\resizebox{0.85\hsize}{!}{\includegraphics[angle=-90]{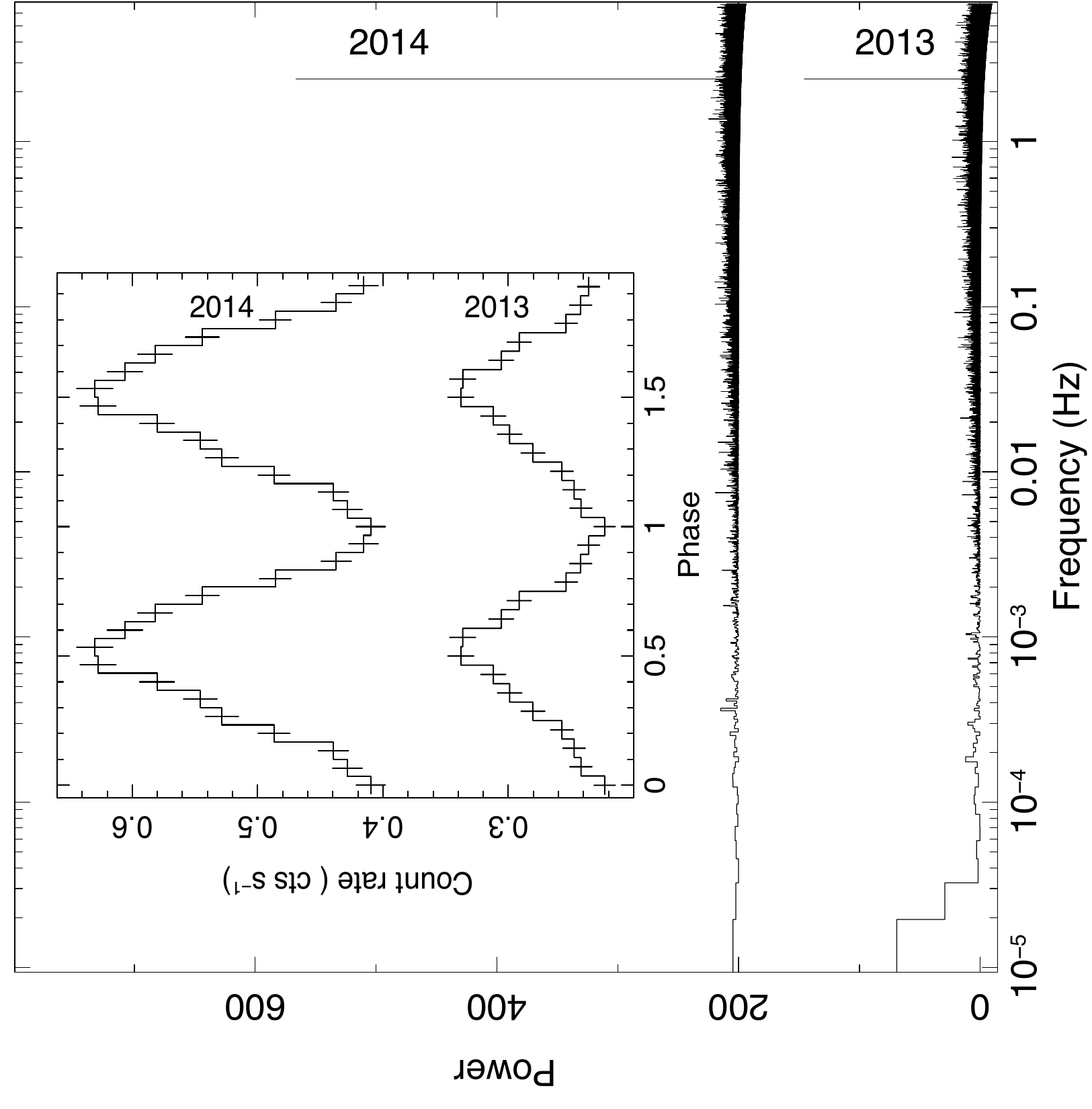}}
\caption{\label{fft_efold} Fourier power spectra for the November 2013 (bottom) and December 2014 (top; shifted by +200 in power) pn data (0.1--12~keV; $\Delta t\simeq73$~ms, 524,288 frequencies). The prominent peaks at about 2.4~Hz  correspond to the 0.42-s signal. The inset shows the background subtracted light curves of the two data sets  folded to their best periods (see Table\,\ref{tabtime}.)}
\end{figure}

\subsection{Discovery of the period and timing analysis}\label{timing}

A periodic signal at about 0.42~s was first detected in the 0.1-12\,keV pn data set 0748390901 (2014 December) at a confidence level {\rc of about 13$\sigma$} by the power spectrum peak detection algorithm of the automatic analysis (\citealt{israel96}; see Fig.\,\ref{fft_efold} and Table\,\ref{tabtime}). By a phase fitting analysis, we honed the period at $0.4183891\pm0.0000001$~s (with uncertainty at 1$\sigma$ confidence level, c.l.). No significant first period derivative was found, with 3$\sigma$ limits of about $\pm$5$\times$$10^{-11}$~s~s$^{-1}$. Pulsations at about 0.42~s were also detected in the pn data of Obs. 069376040, carried out one year before (2013 November). The refined period value was $0.4197119\pm0.0000002$~s. Also in this data set, no first period derivative was found (with 3$\sigma$ limits of about $\pm$$10^{-10}$~s~s$^{-1}$), but  the difference between the two measures implies a long-term average (`secular') period derivative $\dot{P}_{\mathrm{sec}}= -(4.031\pm0.004)\times10^{-11}$~s~s$^{-1}$. 

\begin{figure}
\centering
\resizebox{0.65\hsize}{!}{\includegraphics[angle=-90]{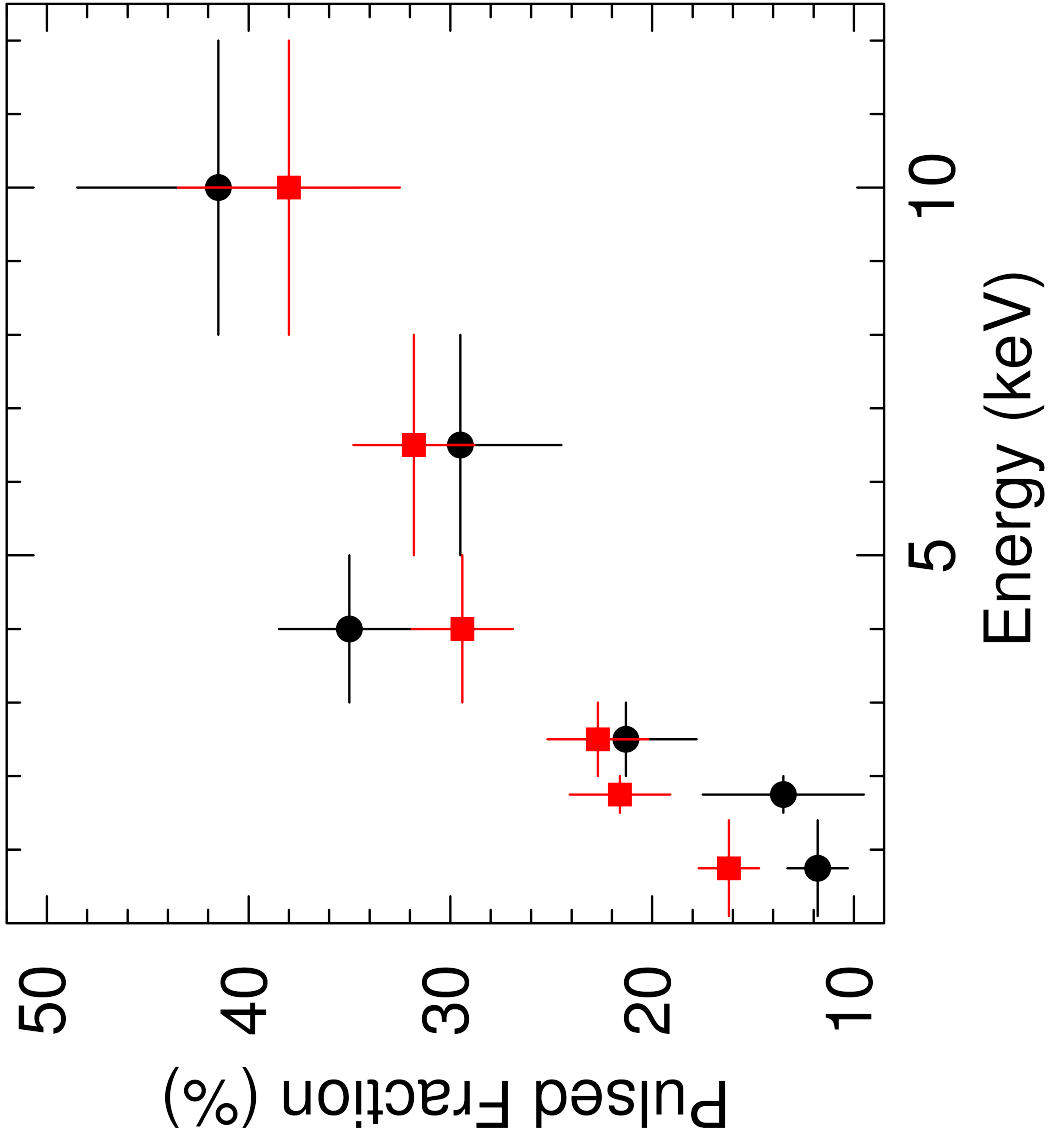}}
\caption{\label{lightcurve} Background subtracted pulsed fractions of the 0.42~s signal as a function of energy for the 2013 (circles) and 2014 (squares) pn data. }
\end{figure}

The 0.1-12\,keV  pulse profiles are almost sinusoidal (single-peaked), with an averaged pulsed fraction of $18\pm1\%$ and $22\pm1\%$ for the 2013 and 2014 observation, respectively (see inset in Fig.\,\ref{fft_efold}).  The pulsed fraction is increasing from 10--20\% below 1.5~keV up to about 40\% above 8~keV (see Fig.\,\ref{lightcurve}).  

\begin{table}
\centering \caption{Main properties of the \src\ pulsar.}\label{tabtime}
\begin{tabular}{lcc}
\hline
Epoch (MJD TDB) & 56621.0&  57001.0\\
$P$ (s) & 0.4197119(2) & 0.4183891(1)\\
$\nu$ (Hz) & 2.382586(1) & 2.3901207(6)  \\
$|\dot{P}|$ (10$^{-11}$~s~s$^{-1}$) & $<$10 & $<$5\\
$\dot{P}_{\mathrm{sec}}$ (10$^{-11}$~s~s$^{-1}$) & \multicolumn{2}{c}{$-$4.031(4)} \\
Pulsed fraction (\%)$^{a}$ & 18(1)& 22(1)\\
$F_X^{0.3-10}$ (\flux) &$1.1(1)$$\times$$10^{-12}$ & $2.7(1)$$\times$$10^{-12}$\\
$L_X^{0.3-10}$ (\lum) & $2.1(2)$$\times$$10^{39}$ & $5.0(2)$$\times$$10^{39}$\\
\hline 
\end{tabular}
\begin{list}{}{}
\item[$^{a}$] Pulsed fraction defined as the semi-amplitude of the sinusoid divided by the average source count rate in the 0.1--12~keV range.
\end{list}
\end{table}

\begin{figure}
\centering
\vspace{3mm}
\resizebox{0.83\hsize}{!}{\includegraphics[angle=0]{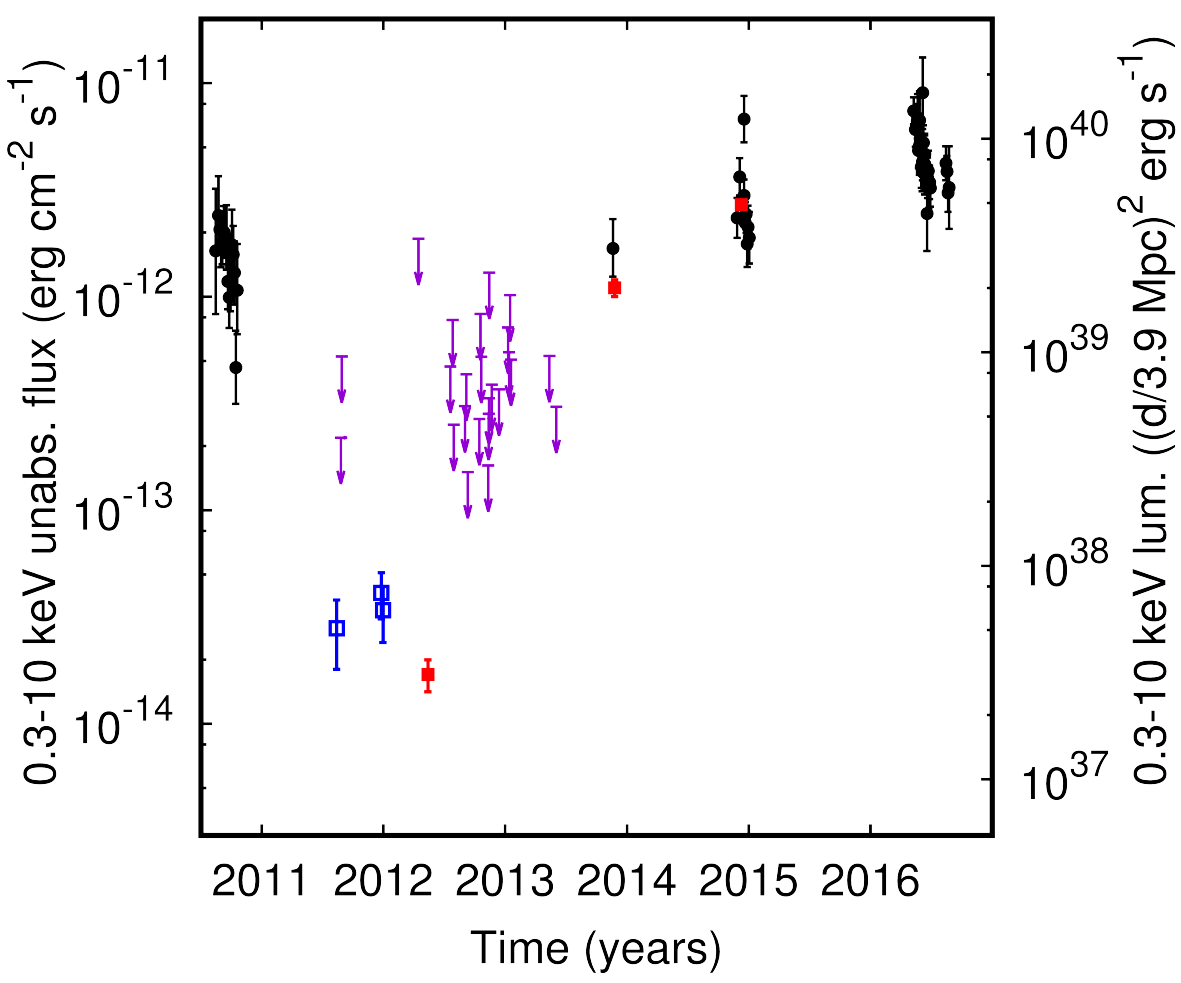}}
\caption{\label{lcurve} Long term 0.3--10 keV flux {\rc (left axis) and isotropic luminosity (right axis)} evolution of \src\ as observed by \swift\  (filled circles), \xmm\ ({\rc filled squares) and \cxo\ (empty squares)}. Arrows mark the 3$\sigma$ upper limits derived from individual \swift\ ({\rc purple})  observations. }
\end{figure}

\subsection{Spectral analysis and long-term variability}\label{spectroscopy}

The spectral fitting was performed in 0.3--10~keV using \textsc{xspec} v.12.8; the abundances used are those of \citet*{wilms00}. 
Different spectral models have been used in literature to fit the 2013 \xmm\ spectrum of \src\ \citep{motch14}. Bearing in mind the nature of the compact object, we simply notice that a good fit  can be also obtained assuming an empirical model often used for accreting X-ray pulsars in our Galaxy: an absorbed power-law with a high energy cut-off (the multiplicative component \textsc{highecut} in \textsc{xspec}) plus, sometimes, a  soft thermal component at low energies, which we modelled with a blackbody (BB). A similar model has been successfully tested on the X-ray spectra of a sample of ULXs with broadband XMM+NuSTAR data by Pintore et al. (2016, in preparation). We note that, regardless of the model used, the fluxes and luminosities derived do not change significantly. The model \textsc{phabs[highecut*(powerlaw+bbodyrad)]} gave the following parameters (we assumed that the two datasets have the same absorption):  $N_{\rm H}=(9.60\pm0.01)\times10^{20}$ cm$^{-2}$,   $\Gamma^{2013}=1.2\pm0.1$ and $\Gamma^{2014}=1.14\pm0.06$, $E_{\mathrm{cut}}^{2013}=5.5^{+0.8}_{-0.5}$~keV and $E_{\mathrm{cut}}^{2014}=6.5^{+0.4}_{-0.6}$~keV, $E_{\mathrm{fold}}^{2013}=5.0\pm1.7$~keV and  $E_{\mathrm{fold}}^{2014}=4.6^{+1.5}_{-0.9}$~keV, $kT_{\mathrm{BB}}^{2013}=0.18\pm0.02$ keV and $kT_{\mathrm{BB}}^{2014}=0.23\pm0.04$ keV, $R_{\mathrm{BB}}^{2013}=(1.2\pm0.3)\times10^3$~km and $R_{\mathrm{BB}}^{2014}=(0.7\pm0.1)\times10^3$~km (90\% c.l. uncertainties are reported; see also Table\,\ref{tabtime} for fluxes and luminosities). We note that the size of the black body is of the order of the corotation radius of the pulsar (see below). The probabilities, as inferred through the Fisher test (F-test), that the inclusion of the \textsc{highecut} component (with respect to the power-law alone) and of the blackbody component (with respect to \textsc{highecut$\ast$powerlaw}), are not needed are 7$\times$10$^{-17}$ and 9$\times$10$^{-12}$, respectively. 
The reduced $\chi^2$ for the simultaneous fit of the two data sets is 1.13 for 373 degrees of freedom (dof). {\rcb For the 2011 \cxo\ spectra we considered a simple absorbed power-law model due to poor statistics, and compared the results with those of the brightest \cxo\ (2003) and \xmm\ (2014) observations (adopting the same model and keeping the same value for  $N_{\rm H}$). We obtained the following value for the photon index:  $\Gamma^{2011}$=1.5(2),  $\Gamma^{2003}$=1.21(3) and $\Gamma^{2014}$=1.29(2) at 90\% c.l., suggesting a moderate steepening of the spectrum at lower fluxes.}

The data from the \swift\ monitoring proved useful to study the long-term variability of \src. We fit {\rc simultaneously} the spectra in the 0.3--10~keV band, using the Cash (C) statistics and  {\rc after binning energy channels so as to have at least one count per bin.} {\rc We assumed} an absorbed power law model, forcing the absorption column to take the same {\rc (free)} value in all the observations. The C statistics we obtained was 3906 for 4595 dof. We measured an absorption column density of $N_{\rm H}=(6.7\pm0.2)\times10^{20}$~cm$^{-2}$. The average value of the index of the power law was $<$$\Gamma$$>=1.03$, with a standard deviation of 0.18. The observed 0.3--10~keV unabsorbed fluxes are plotted in Fig.\,\ref{lcurve}; the right scale of the plot represents the isotropic luminosity in the same energy band. When the source was not significantly detected, we set an upper limit on the 0.3--10~keV count rate at 3$\sigma$ confidence level following \citet{gehrels86}. Then, we converted the count rate into a flux estimate with WebPIMMS,\footnote{See \mbox{https://heasarc.gsfc.nasa.gov/cgi-bin/Tools/w3pimms/w3pimms.pl}.} assuming that the spectrum is described by an absorbed power law with absorption column $N_{\rm H}=5.7\times 10^{20}$~cm$^{-2}$ and photon index $\Gamma=1$ (see arrows in Fig.\,\ref{lcurve}).

\src\ was detected in X-ray outburst during observations performed in 2010, and from late 2013 to 2016. The maximum and minimum observed 0.3--10~keV flux are $9^{+4}_{-3}\times10^{-12}$ and $(0.5\pm0.2)\times10^{-12}$~\flux\ observed on 2016 June 6 and 2010 October 15, respectively. These fluxes translate into 0.3--10~keV isotropic luminosity of $\sim$$1.6\times10^{40}$ and $\sim$$9\times10^{38}$~\lum. {\rc A bolometric correction factor of ∼1.25 is obtained if it is assumed that the spectrum is cut off at $E_{\rm cut}=5.5$~ keV with a folding energy of $E_{\rm fold}=5$~keV.}

\begin{figure}
\centering
\resizebox{.91\hsize}{!}{\includegraphics[angle=-90]{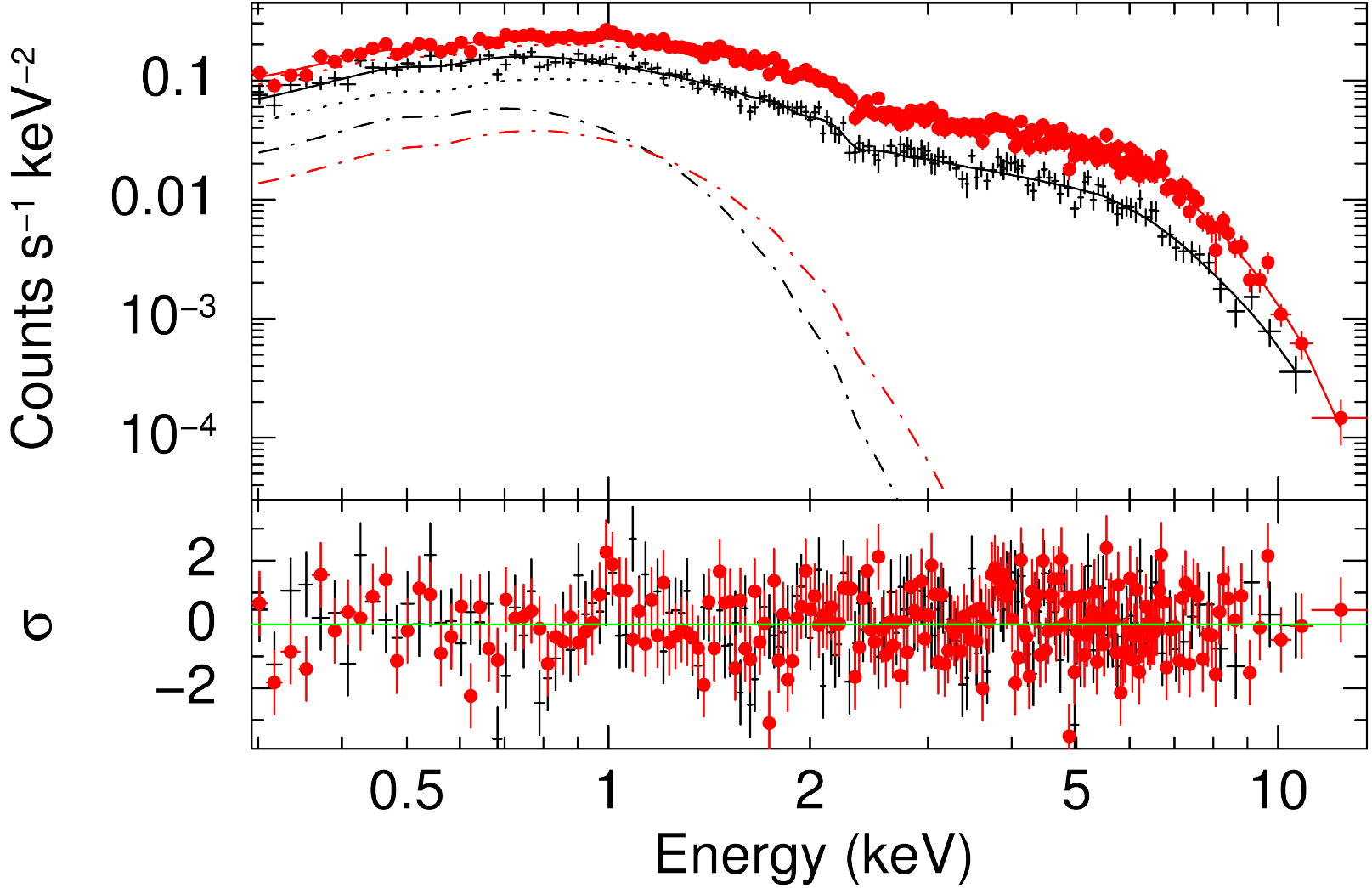}}
\caption{\label{spec} \xmm\ pn energy spectra of \src\ fit with the {\rc blackbody (dot-dashed line)  plus powerlaw (dotted line) model} described in Sect.\,\ref{spectroscopy} for the 2013 (crosses) and 2014 (filled circles). The residuals, in unit of $\sigma$, are also plotted (lower panel).}
\end{figure}

\section{Discussion}\label{discussion}

\citet{motch14} found that the orbital period of NGC 7793 P13 is 64~d and that the properties of the optical counterpart are consistent with those of a B9Ia supergiant companion with mass in the range $M_2 = 18$--23~M$_\odot$ and radius of  \mbox{$R_2 = 50$--60~R$_\odot$}. The same authors assume that the star fills its Roche lobe, since the stellar wind from a B9Ia supergiant cannot provide the accretion rate needed to produce the maximum observed X-ray luminosity ($10^{19}$--$10^{20}$~g~s$^{-1}$, see below). All acceptable orbital solutions  require a significant eccentricity ($e$=0.27--0.41). From this constraint, \citet{motch14} conclude that, at periastron, the supergiant can fill its Roche lobe for a BH companion with a mass 3.4~M$_{\odot} < $~M$_{\mathrm{BH}} < 15$~M$_{\odot}$, with the upper limit being fixed by the requirement that the Roche lobe is not too small to accommodate the star.

The discovery of a pulsar in NGC 7793 P13 allows us to place an independent and even tighter constraint on the orbital eccentricity of the system. Assuming a mass $M_1 = 1.4$~M$_{\odot}$ for the NS, the Roche lobe of the companion is bigger than that for a BH. Therefore, even a B9Ia supergiant cannot fill it unless the eccentricity is $e$=0.46--0.55, so that, at periastron, the separation is sufficiently small to allow for a contact phase. 

Following \citet{motch14}, we assume that the mass transfer in the system proceeds on a thermal timescale $t_{\mathrm{th}}$: $
 t_{\mathrm{th}} =  2.4\times 10^5 (M_2/ 20~\mathrm{M}_\odot)^2 \times  
        (R_2/50~\mathrm{R}_\odot)^{-1} (L_2/10^4 \mathrm{L}_\odot)^{-1} \, {\rm yr} \, ,
\label{eq1}
$
where $L_2$ is the luminosity of the companion. The mass transfer rate is then:
$
  {\dot M_2} \approx M_2 / t_{\mathrm{th}} =  3.7 \times 10^{21} (M_2/ 20~\mathrm{M}_\odot)^{-1} \times 
  (R_2/50 ~\mathrm{R}_\odot) (L_2/10^4 ~\mathrm{L}_\odot) ~{\rm g~s^{-1}}.
$
Owing to the large mass ratio ($q=M_2/M_1\gg1$), the evolution is expected to be non-conservative. 
Part of the mass is likely to be removed from the binary through hydrodynamical instabilities, although the system may  possibly be stabilized by the significant mass loss of the companion \citep{fragos15}. Even considering all mass losses, ${\dot M_2}$ is so high to easily account for the accretion rate ${\dot M}$ implied by the maximum observed X-ray luminosity:
$
  L_{\mathrm{max}} \approx b \cdot 10^{40}~\mathrm{erg~s^{-1}},~  
  {\dot M} = L_{\mathrm{max}}/(\eta c^2) \approx 10^{20} b \, (0.1/\eta) \, {\rm g~s^{-1}} 
$,
where $b$ is the beaming factor and $\eta$ the accretion efficiency (see also below).

\src\ displayed a factor of $\sim${\rc 8} 
flux variation {\rc in the high state (above $\sim$2$\times$10$^{39}$\lum; see Fig.\,\ref{lcurve})}, attaining a maximum isotropic luminosity of 
$L_{\mathrm{iso}}^{\mathrm{max}}\sim 1.6\times10^{40}$~\lum, about 100 times higher than 
the Eddington limit.  We note that the $\sim$0.42~s pulsations 
were observed at the top and close to the bottom of this range, implying that accretion onto 
the NS took place over the entire interval of variation. 

\begin{figure}
\vspace{-9mm}
\resizebox{1.18\hsize}{!}{\includegraphics[angle=0]{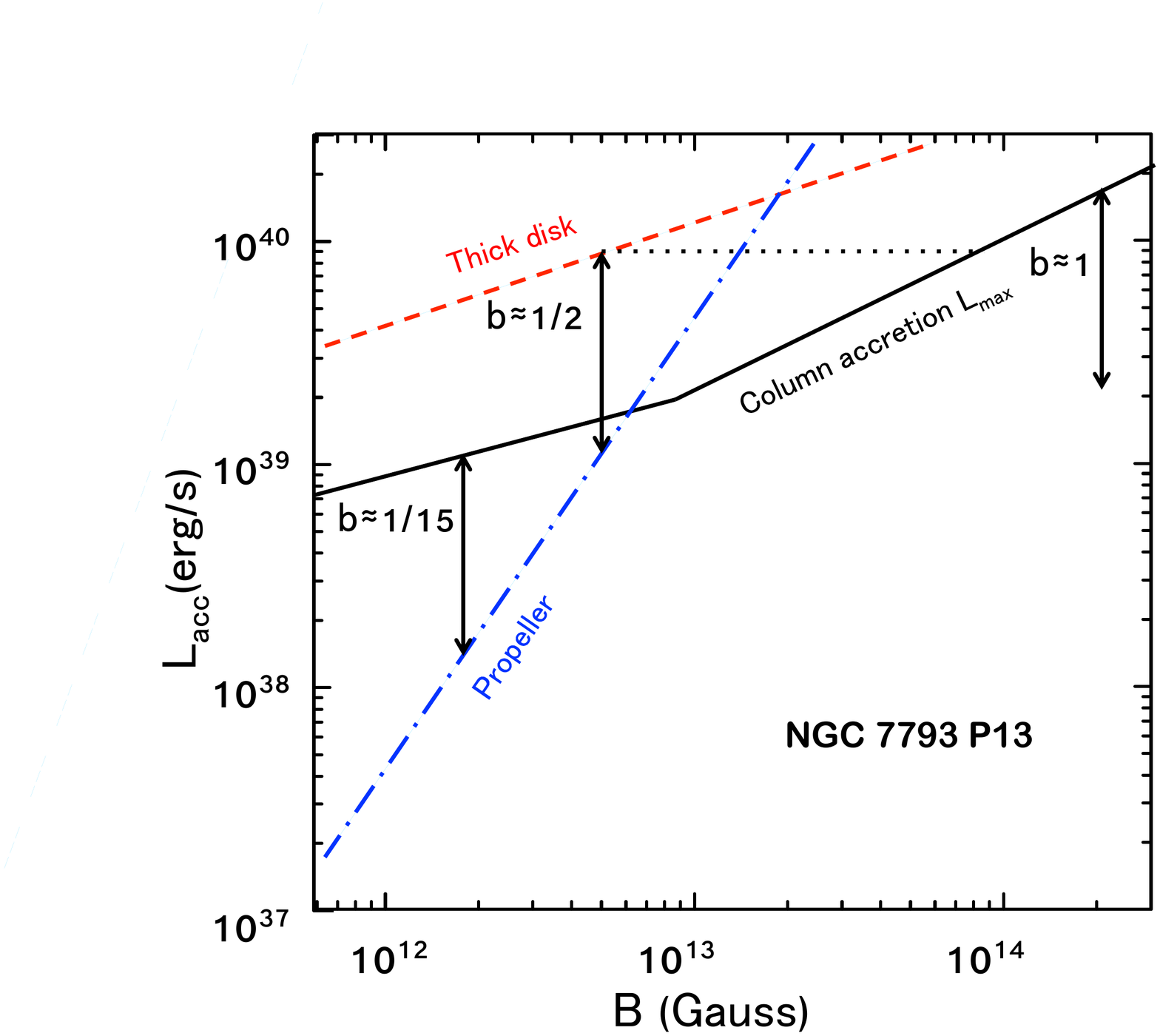}}
\vspace{-6mm}
\caption{\label{Bfield} Accretion luminosity versus surface magnetic field constraints for \src. The solid line gives the maximum luminosity that can be produced by column-accretion onto the NS magnetic poles. The dashed line is the limit above which the energy released in the accretion  disk exceeds the Eddington limit and disk thickens.  Accretion is inhibited below the dot-dashed line, as the NS enters the propeller regime.  Double-arrowed segments show the factor of $\sim${\rc 8}  flux variation displayed by the source  when pulsations were detected.
Different segments are shifted by the inverse of the beaming factor  $b^{-1} = L_{\mathrm{iso}}/L_{\mathrm{acc}}$. A value of $b\sim$1/2 is found to be in agreement with the observed source properties (see text): this solution implies a dipole magnetic field of $B$$\sim${\rc 5} 
$\times$$10^{12}$\,G and a multipolar field $B$$>$${\rc 8} 
$$\times$$10^{13}$\,G at the base of the accretion column (dotted line).}
\end{figure}

In our discussion here we assume that accretion onto the NS takes place unimpeded (at least) over the above mentioned luminosity variation.  The $\sim -4.0\times10^{-11}$~s~s$^{-1}$ period derivative inferred from the two one-year-apart observations during which pulsations were detected,
is virtually unaffected by orbital Doppler shift given that the two \xmm\ pointings are almost at the same orbital phase (assuming P$_{orb}=$ 63.52\,d; \citealt{motch14}).

NSs may attain accretion luminosities exceeding the Eddington limit by orders of magnitude if their surface magnetic field ($B$) is very high, so that electron scattering cross sections for extraordinary mode photons below the cyclotron energy $E_{\mathrm{c}} \sim 12 (B/10^{12}$\,G)~keV is much lower than the Thomson cross section. \cite{mushtukov15}  show that column-accretion onto a $>10^{15}$~G magnetic pole may give rise to a luminosity of $L\sim10^{41}$~\lum. However, for a magnetic NS to accrete at a very high rate, other conditions must be met. First, the accretion flow outside the magnetosphere must take place through a disk that remains geometrically thin ({\it i.e.} height/radius ratio $<$1), such that the bulk of the flux emitted close to the bottom of the accretion column can escape.
This translates into the condition that the accretion energy released in the disk down to the 
magnetospheric radius $r_{\mathrm{m}}$ is sub-Eddington (dashed line in Fig.\,\ref{Bfield}).
An additional condition is that the NS angular velocity is smaller than the (Keplerian) angular velocity of the disk at $r_{\mathrm{m}}$, so that the drag exerted by the rotating magnetic field lines as matter enters the magnetosphere is weaker than gravity and matter can accrete onto the surface. This is equivalent to requiring that  $r_{\mathrm{m}} < r_{\mathrm{cor}} $, where $r_{\mathrm{cor}} = \left(\frac{G M P^2}{4 \pi^2}\right)^{1/3}$  is the corotation radius \citep{illarionov75,stella86}. When $r_{\mathrm{m}} > r_{\mathrm{cor}}$, centrifugal forces at $r_{\mathrm{m}}$ exceed gravity and only little accretion, if any, can take place when the so-called propeller regime ensues (dot-dashed line in Fig.\,\ref{Bfield}). For \src\ to emit isotropically a maximum luminosity of $L^{\mathrm{max}}_{\mathrm{iso}}\sim 1.6 \times 10^{40}$~\lum\ according to the model of \cite{mushtukov15} the NS surface dipolar magnetic should be at least $B\sim 2 \times 10^{14}$~G.  However, for such value of $B$ and $P\sim0.42$~s, accretion would be inhibited by magnetospheric drag and the NS would be deep in the propeller regime.

Therefore we relax the assumption of isotropy, and consider that the 
NS emission is beamed by a factor $b < 1$. In this case the isotropic equivalent luminosity is $L_{\mathrm{iso}} = L_{\mathrm{acc}}/b$ and the accretion luminosity $L_{\mathrm{acc}} = GM\dot M/R$ is reduced correspondingly  (here $R$ and $M$ are the NS radius and mass). 
We assume that the minimum (detected) isotropic luminosity of $L^{\mathrm{min}}_{\mathrm{iso}} \sim L^{\mathrm{max}}_{\mathrm{iso}}/{\rc 8} 
$ marks the onset of the transition  from accretion to the propeller phase ({\it i.e.} $r_{\mathrm{m}} = r_{\mathrm{cor}}$) and, at the same time, require that the surface magnetic field is high enough to attain the observed luminosity range (solid line in Fig.\,\ref{Bfield}). A surface dipole field of {\rc $B\sim 2\times 10^{12}$~G}
and a maximum accretion luminosity of $L_{\mathrm{acc}} \sim 10^{39}$~\lum\ are obtained, corresponding to beaming factor of $b\sim 1/15$ (note that for these parameters the disk remains geometrically thin). For a time-averaged accretion rate 
of $\dot M \sim 3 \times 10^{18}$~g~s$^{-1}$, as implied by this solution, we estimate the corresponding  maximum spin-up rate by imposing that the matter accreting onto the NS carries the Keplerian angular momentum at $r_{\mathrm{cor}}$. This gives a maximum $\dot P = \dot{M}\ r_{\mathrm{cor}}^2P/I \sim - 1 \times 10^{-11}$~s~s$^{-1}$, four times smaller than the secular $\dot P$ derived from the data. 

In order to ease this problem, and by analogy with the {\rc case of NGC\,5907 ULX-1 \citep{israel16b}},   we consider the possibility that close to the NS surface (and thus the base of the accretion column) the magnetic field is dominated by higher than dipole magnetic multipoles. Close to the magnetospheric radius ($r_{\mathrm{m}} \sim 10^8$~cm) the field is virtually the dipolar by virtue of its less steep radial dependence. This is done by analogy with the case of magnetars \citep{thompson95,tiengo13}.
The conditions that the accretion disk is thin for $L^{\mathrm{max}}_{\mathrm{iso}} = L^{\mathrm{max}}_{\mathrm{acc}}/b$ and that 
the NS is in the accretion regime for $L^{\mathrm{min}}_{\mathrm{iso}} = L^{\mathrm{min}}_{\mathrm{acc}}/b$ depend on the B-field strength at $r_{\mathrm{m}}$, where only the dipole component matters. Both conditions are satisfied for $B$ of {\rc $\sim 5 \times 10^{12}$~G}
and $b \sim 1/2$. A (multipolar) $B$ $>8 \times 10^{13}$~G
at the base of the  accretion column would be required to give rise to corresponding maximum accretion luminosity of {\rc $L^{\mathrm{max}}_{\mathrm{acc}} =9\times 10^{39}$ \lum (dotted line in Fig.\,\ref{Bfield}). 
A maximum spin-up of $\dot P \sim -7 \times 10^{-11}$~s~s$^{-1}$ is derived in this case (owing to the higher time-averaged accretion rate resulting from $b \sim 1/2$), consistent with the value inferred from the observations.  For a $\sim 0.42$~s
spin period we expect that the isotropic luminosity in the propeller regime is $< L^{\mathrm{min}}_{\mathrm{iso}}/90 \sim 2\times10^{37}$~\lum \citep{corbet97}, a level which is slightly below the values inferred during the 2011 \cxo\ and 2012 \xmm\ observations, to within uncertainties.}

The discovery of three PULXs  
previously classified as stellar mass BH based on their spectral properties, strongly suggests that this class might be more numerous than suspected so far, and that other know ULXs might host an accreting NS. The large first period derivative, the intermittance of the pulsations and their relatively small pulsed fraction make their detection a difficult task.

\vspace{-.5cm}

\section*{Acknowledgements} 
EXTraS is funded from the EU's Seventh Framework Programme under grant agreement n. 607452. 
PE acknowledges funding in the framework of the NWO Vidi award A.2320.0076. LZ and GLI acknowledges funding from the ASI - INAF contract NuSTAR I/037/12/0. AP acknowledges support via an EU Marie Sklodowska-Curie Individual Fellowship under contract No. 660657-TMSP-H2020-MSCA-IF-2014.
{\rcb After submission we become aware of the manuscript by F\"uerst et al. (2016, arXiv:1609.07129) which confirmed the discovery.}  

\vspace{-.5cm}

\bibliographystyle{mn2e}
\bibliography{biblio}

\bsp
\label{lastpage}

\end{document}